\documentclass[fleqn,a4paper,10pt,twocolumn]{scrartcl}
\usepackage{graphicx,amsmath}
\usepackage[footnotesize]{caption2}

\newcommand{\dderiv}[3][]{\frac{\mathrm{d}^{#1}{#2}}{\mathrm{d}{#3}^{#1}}}
\newcommand{\Ec}{E_\mathrm{C}}
\newcommand{\Eceff}{E_{\mathrm{C,eff}}}
 
\newcommand{\ff}[1]{\tfrac{1}{#1}}
\newcommand{\ffi}[1]{\frac{1}{#1}} 

\title{Coulomb blockade thermometry using a two-dimensional array of tunnel
junctions}
\author{Tobias Bergsten, Tord Claeson and Per Delsing\\
\small\textit{Department of Microelectronics and Nanoscience, Chalmers
University of Technology}\\
\small\textit{and G\"oteborg University, SE-412 96 G\"oteborg, Su\`ede}}
\date{}

\begin{document}
\twocolumn[
\maketitle
\vspace{-10mm}

\small\quotation\centering
\begin{minipage}{142mm}
We have measured current-voltage characteristics of two-dimen\-sional 
arrays of small tunnel junctions at temperatures from 1.5\,K to 4.2\,K. 
This corresponds to thermal energies larger than the charging energy.  We 
show that 2D-arrays can be used as primary thermometers in the same 
way as 1D-arrays, and even have some advantages over 
1D-arrays.  We have carried out Monte Carlo simulations, which 
agree with our experimental results.
\end{minipage}\\
\vspace{10mm}
] 

\section{Introduction}
Thermometers can be divided into two categories: primary and 
secondary ones.  The former do not have to be calibrated, but are 
generally slow and expensive, and are therefore mainly used for 
calibration purposes.  Secondary thermometers are, on the other hand, 
usually cheap and fast, but they need to be calibrated against a known 
temperature scale, i.e.  a primary thermometer or a scale which can be 
traced to one.  The vast majority of thermometers used for routine 
measurements are secondary thermometers.

\begin{figure}
\center
\includegraphics[width=0.9\columnwidth]{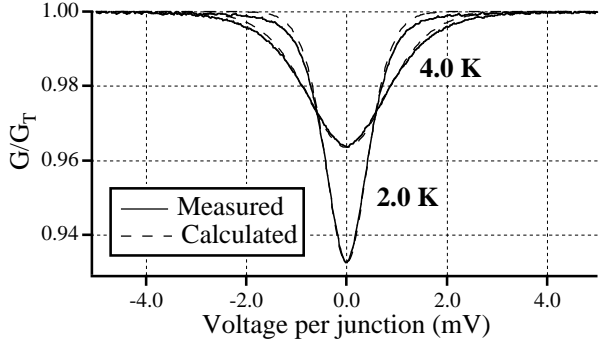}
\caption{The differential conductance as a function of bias voltage shows a 
bell-shaped dip around zero voltage.  These measurements were made on a 
$256\times 256$ junction 2D-array with a resistance of 17\,k$\Omega$ 
and an effective capacitance of 2\,fF.  The calculated curves were 
calculated with higher order corrections \cite{farhang}.}
\label{cvgraf}
\end{figure}

Recently, a new primary thermometry method, the Coulomb blockade 
thermometry (CBT), was demonst\-rated by Pekola \textit{et al.}\cite{pekolaprl} It 
is based on the charging effects in one-dimensional (1D) arrays of 
very small tunnel junctions.  When the charging energy of the 
junctions, $\Ec=e^2/2C$, is greater than or of the order of the thermal energy,
$k_BT$, the finite charge of the electron gives rise to a decreased 
differential electrical conductivity at low voltages ($C$ is the tunnel 
junction capacitance and $T$ is the temperature).  This is called the Coulomb 
blockade of tunnelling \cite{averin,grabert}.  The Coulomb blockade 
shows up as a bell-shaped dip (figure \ref{cvgraf}) when the 
differential conductance, $\dderiv{I}{V}$, is plotted against the bias 
voltage, $V$.  The depth of this dip is inversely proportional and the 
half-width is directly proportional to the temperature.  
Interestingly, the half-width depends only on the temperature, $T$, 
and natural constants, which makes it useful for primary thermometry.  
This method has the advantage over other primary methods used at cryogenic 
temperatures in that it is both faster and cheaper, furthermore the 
physical size of the thermometer can be made less than a mm.  The 
depth of the conductance dip depends on the tunnel junction 
capacitace, $C$, as well.  Pekola \textit{et al.} have demonstrated 
that this method can give an accuracy in temperature of at least 0.5\% 
and they have used it successfully in the range from 20\,mK to 30\,K, 
more than three orders of magnitude, which is a large range for a 
primary thermometer.

In this paper we show experimental data for a CBT consisting of a 
two-dimensional (2D) array of tunnel junctions, rather than the 
1D-arrays demonstrated by the group of Pekola \textit{et al.}
\cite{pekolaprl,farhang,kaupprb} One practical problem with 1D-arrays of 
tunnel junctions is that they are dependent on every junction to 
function properly, like a chain is only as strong as its weakest link.  
So if one junction is damaged, ages rapidly or simply is bad from the 
beginning, the array might be unusable or at least its performance 
will be degraded.  One way to reduce this degradation due to bad junctions is 
to connect several arrays in parallel, thus reducing the error to the 
order of $1/M$, where $M$ is the number of arrays.  Even better is to 
make a 2D-array, where the metal islands are connected with tunnel 
junctions both in parallel and in series.  The current would then 
simply pass around any broken junctions.  This reduces the error due 
to a broken junction to the order $1/MN$, where $M$ and $N$ are the 
number of junctions in parallel and in series respectively.  (The 
resistance of an array of resistors increases with a factor of 
approximately $2/MN$ if one resistor is removed.)

We have measured current-voltage characteristics on a 2D array of 
tunnel junctions at temperatures from 1.5\,K to 4.2\,K in a pumped 
$^4$He cryostat.  Using the formulas derived for 1D-arrays we have 
calculated the temperatures from these IV-curves and compared them 
with temperatures calculated from the pressure of the He gas in the 
cryostat.  The correspondence is very good.

We have also made Monte-Carlo simulations of 1D- and 2D-arrays and these 
simulations show that the tunnel junction capacitance $C$ which is 
used in the equations for 1D-arrays should be replaced in the 2D case 
by an effective capacitance $C_\mathrm{eff}$ which is higher than the real 
capacitance.

\section{Theory}
The differential conductance $G$ of a 1D-array of $N$ junctions has been 
calculated theoretically \cite{pekolaprl} to first order in the limit 
$\Ec\ll k_BT$.  For arrays where all junctions have the same 
resistance and capacitance and negligible capacitance to ground, the 
result is

\begin{equation}\label{ggt}
    \begin{split}
    \frac{G}{G_T}&=1-u_Ng(v_N),\\ 
    g(x)&=\frac{x\sinh(x)-4\sinh^2\frac{x}{2}}{8\sinh^4\frac{x}{2}}
    \end{split}
\end{equation}
where the parameters $u_N$ and $v_N$ are defined as
\begin{equation}\label{un}
    u_N=2\frac{N-1}{N}\frac{\Ec}{k_BT}=2\frac{\Eceff}{k_BT},
\end{equation}
\begin{equation}
    v_N=\frac{eV}{Nk_BT}.
\end{equation}

$G_T$ is the differential conductance at voltages far above the 
Coulomb blockade.  The effective charging energy, $\Eceff$ is defined 
similarly to the usual charging energy, except the physical capacitance 
is replaced with the effective capacitance (which in a 1D-array with 
negligible capacitance to ground is $C_\mathrm{eff}=\frac{N}{N-1}C$).

The half-width $V_{\ffi{2}}$ of the resulting conductance dip depends 
only on $N$ and $T$:

\begin{equation}\label{v12}
    eV_{\ffi{2}}=5.439Nk_BT.
\end{equation}

The depth $\frac{\Delta G}{G_T}=\ff{6}u_N$ of the conductance dip also
depends on the effective capacitance $C_\mathrm{eff}$ of the junctions.

At low temperatures ($k_BT\approx\Ec$) higher order corrections should 
be included \cite{farhang}. This gives a correction 
to the halfwidth in equation (\ref{v12}), which is independent of 
temperature:

\begin{equation}\label{dv}
\begin{split}
    eV_{\ffi{2}} &=5.439Nk_BT\left(1+0.3921\frac{\Delta G}{G_T}\right)\\
                 &=5.439Nk_BT+0.7108 N\Eceff.
\end{split}
\end{equation}

\section{Sample parameters}
The measured array consisted of $256\times 256$ tunnel junctions in a square 
pattern with a side of 200\,$\mu$m and the area of the overlap 
junctions was approximately 150\,nm squared.  The tunnel junctions 
were made of aluminium with a tunnel barrier consisting of aluminium 
oxide, AlO$_x$, fabricated with standard shadow evaporation technique 
\cite{niemeyer,dolan}.  The junctions had a tunnelling resistance of 
17\,k$\Omega$ and a capacitance of 1.1\,fF which means $\Ec/k_B$ was 
0.85\,K.  The capacitance was calculated from the offset voltage, 
$V_\mathrm{off}$=9.0\,mV, using the offset analysis method described 
in ref \cite{wahlgren}.  We have furthermore assumed that the offset 
voltage of a 2D-array of tunnel junctions is $V_\mathrm{off}=eN/4C$ 
\cite{bakhvalov}.

\begin{figure}
\center
\includegraphics[width=\columnwidth]{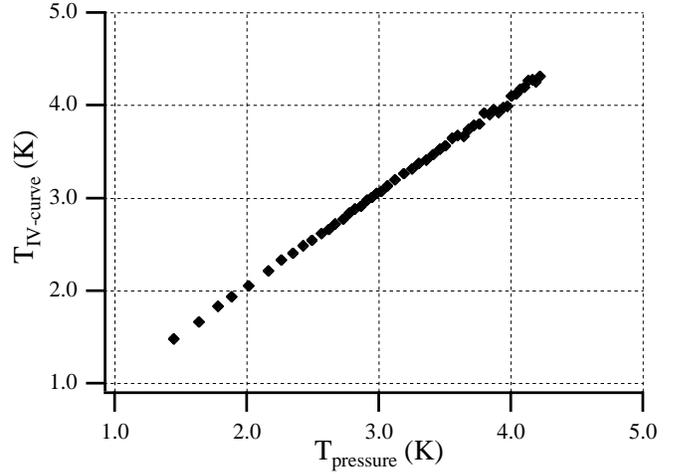}
\caption{Comparison between the temperature calculated from the 
half-width of the Coulomb blockade in the IV-curve and the 
temperature calculated from the $^4$He equilibrium pressure.}
\label{ttgraf}
\end{figure}

\section{Experimental setup}
The measurements were carried out in a pumped $^4$He cryostat which 
has a temperature range from 4.2\,K down to approximately 1.2\,K.  A 
custom built biasing and amplification box fed the source signal to 
the sample and amplified the current and voltage signals, which were 
measured with two Keithley 2000 multimeters.  A computer controlled 
Keithley 213 voltage source was used for the voltage sweep.  The 
$^4$He bath pressure was measured by a Wallace \& Tiernan FA129 analog 
pressure meter which was connected directly to the pumped helium space.

\section{Results}
We measured IV-curves at 50 different temperature points and 
differentiated them numerically.  Using equation \ref{dv} we 
calculated the temperature and compared it with the temperature 
calculated from the $^4$He equilibrium pressure in the He bath, using 
the definition of ITS-90 \cite{its90}.  As can be seen in figure \ref{ttgraf} the 
two temperatures are almost the same, but not quite.  The temperature 
from the IV-curve is consistently slightly higher than the temperature 
from the $^4$He pressure.

\begin{figure}
\center
\includegraphics[width=\columnwidth]{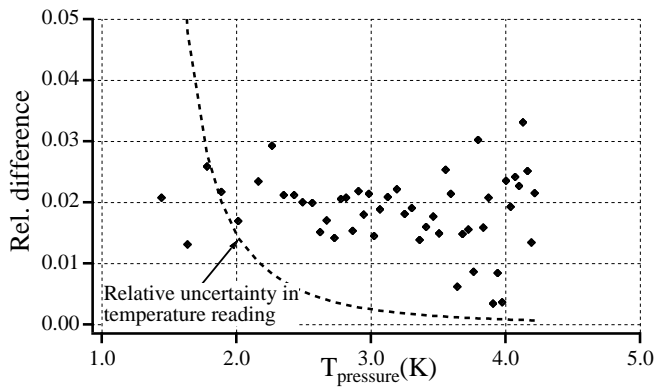}
\caption{The relative difference between the temperature calculated
from the IV-curve and from the $^4$He pressure.  There is a systematic 
difference of around 2\%.  As a comparison the uncertainty from the pressure 
reading is included, but it can't account for all the difference in 
temperature.}
\label{diffgraf}
\end{figure}

The difference is clearer if we plot the relative difference, 
\textit{i.e.} $(T_{IV}-T_{pres})/T_{pres}$ (figure \ref{diffgraf}).  
We see that there is a positive difference of 1-3\% in all the 
measurements.  The uncertainty in the pressure readings was estimated 
to about 1 Torr and the resulting temperature uncertainty is also 
shown in figure \ref{diffgraf}.  While it may account for some of 
the spread, it can not explain the systematic difference of about 2\%.  One 
possible explaination could be that the IV-curves were measured quite 
rapidly one after the other, and perhaps the sample did not quite cool down 
at each temperature before the measurement was made.

We have made Monte Carlo simulations of one- and two-dimensional 
tunnel junction arrays with a program package called \textsc{simon} 
\cite{wasshuber} which is intended for simulations of circuits with very 
small tunnel junctions.  Because this method turned out to be very 
time-consuming, we settled for quite small arrays, only three 
junctions in series and two in parallel.  Both 1D and 
2D arrays behaved according to theory, except that in 
the 2D case the capacitance in equation \ref{un} was not the actual 
capacitance of the tunnel junctions, but an effective capacitance 
$C_\mathrm{eff}$ which was approximately 1.4 times the actual value in the 
$3\times 2$ junction array.  This is not surprising, though, since two 
islands are not only connected via the junction between them, but also 
via the other islands in the array.  In an infinite 2D array the 
capacitance between two neighbouring islands is exactly twice the 
capacitance of the tunnel junctions, and this is the reason that the 
offset voltage is a factor of two lower in a 2D-array compared to a 
1D-array \cite{bakhvalov}.  By calculating the effective capacitance from 
$\frac{\Delta G}{G_T}$ we get $C_\mathrm{eff}\approx 2.2\,fF$, which 
is twice the capacitance calculated from the offset voltage, as 
expected.

An important property of the 2D-array is that it can be fabricated 
with lower resistance than a 1D-array, even if it contains many 
junctions in series.  This means that the measurement can be done faster 
and the measurement error is lower. When deducing the temperature from 
the half-width of the conductance vs voltage curve, the uncertainty in 
T depends both on the uncertainty in G and in V.  The measurement 
uncertainty can be written

\begin{equation}
    \frac{\delta T}{T}=\frac{4.60\frac{\delta G}{G}}
    {u_N}+\frac{\delta V}{V_{\ffi{2}}},
\end{equation}
\begin{equation}\label{dG}
    \frac{\delta G}{G}=\frac{\delta I}{\Delta I}+\frac{\delta V}{\Delta V}+ 
    \frac{u_N}{720}\left(\frac{e\Delta V}{Nk_BT}\right)^2
\end{equation}

\begin{figure}
\center
\includegraphics[width=\columnwidth]{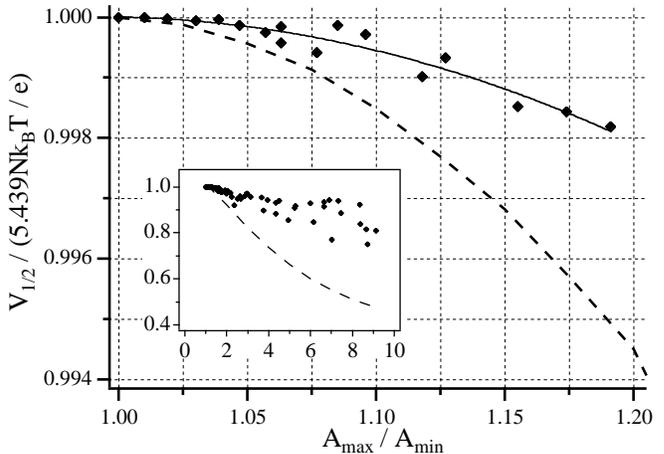}
\caption{The simulated half-width voltage, scaled by the ideal value, 
plotted against $A_{\textrm{max}}/A_{\textrm{min}}$ (dots) for a 
$5\times 5$ junction array.  $A_{\textrm{max}}$ and $A_{\textrm{min}}$ 
are the maximum and minimum junction areas respectively.  The dashed 
lines are the corresponding results for 1D-arrays by Hirvi \textit{et 
al.}\cite{hirvi}, and the solid line is a quadratic fit to the 
simulated values, to compare with the quadratic dependence in the 
1D-array.  It is clear that 2D-arrays are less affected by variations 
in the tunnel junction areas than 1D-arrays, both at moderate 
(big figure) and large (inset) variations.  Larger arrays give 
essentially the same result.}
\label{errorgraf}
\end{figure}

where $\delta I$ and $\delta V$ are the current and voltage 
uncertainties respectively, given by the array resistance and 
temperature, and the amplifier noise.  $\Delta I$ and $\Delta V$ are 
the intervals used for the numerical differentiation of the IV-curves 
or, if lock-in measurements are used, the modulation ampliture. Note that 
$\Delta I=G_{\ffi{2}}\Delta V$ where $G_{\ffi{2}}$ is the differential 
conductance at the half-width voltage.  The last term in eq.  
\ref{dG} is the maximum differentiation error.  Equation \ref{dG} has 
a minimum with respect to $\Delta V$ and we can chose $\Delta V$ (in 
our case approximately 5 \% of $V_{\ffi{2}}$) to minimise the total error:

\begin{equation}
    \frac{\delta 
    T}{T}=0.971\left(\frac{R_\mathrm{T}\frac{\delta I}{M}+\frac{\delta V}{N}}
    {e/C_\mathrm{eff}}\right)^{2/3}+\frac{\delta V}{V_{\ffi{2}}}.
\end{equation}

$R_\mathrm{T}$ is the tunnelling resistance per junction at the 
halv-width voltage.  Note that the uncertainty decreases both with 
increasing $N$ and $M$.

Now we want to compare the measurement uncertainty of a 1D- 
and a 2D-array.  In this example we use a typical 1D-array of 30 
junctions (typical because this is the approximate size used by Pekola 
\textit{et al.}), each junction with an effective capacitance of 2\,fF 
and a resistance of 20\,k$\Omega$.  For the 2D-array we assume $256\times 256$ 
junctions and the same junction parameters. The temperature is 4\,K.  For 
the measurement we assume AD743 operational amplifiers at the input stage, 
because they have low voltage noise.  The resulting uncertainty is 
$1.2\cdot 10^{-3}$ for the 1D-array and $0.22\cdot 10^{-3}$ for the 
2D-array, \textit{i.e.} five times lower in the 2D case.  One might 
argue that we would get the same effect if we fabricated 256 parallel 
1D-arrays with 256 junctions each, but then we might just as well make 
it 2D, for the sake of robustness, as discussed earlier.

Another important issue is the tolerance to variations in the tunnel 
junction properties.  Since no fabrication process is perfect, small 
variations can always be expected and in general, the smaller the 
structures, the bigger are the relative differences.  Hirvi \textit{et 
al.} have calculated the measurement error which can be expected in a 
1D-array at different inhomogenities \cite{hirvi}.  We have simulated 
a $5\times 5$ junction 2D-array with random distributions of tunnel 
junction areas and plotted the resulting half-width voltage 
$V_{\ffi{2}}$ against the parameter 
$A_{\textrm{max}}/A_{\textrm{min}}$, where $A_{\textrm{max}}$ and 
$A_{\textrm{min}}$ are the maximum and minimum junction areas 
respectively (figure \ref{errorgraf}).  Increasing the number of 
junctions give essentially the same result.  We have assumed that 
$R_{i,j}C_{i,j}=\mbox{\textit{constant}}$ for all junctions, which 
follows from uniform tunnel barriers.  It is clear from the figure 
that the 2D-array is much less affected by variations in the junction 
areas than 1D-arrays.  For example, at a 10\% spread 
($A_{\textrm{max}}/A_{\textrm{min}}=1.10$) the error in a 1D-array is 
$\approx$0.15\%, while in a square 2D-array it is $\approx$0.05\%.  It 
should be noted that even a $2\times N$-array is significantly better 
than a 1D-array in this respect, at 10\% spread the error is 
$\approx$0.08\%.

\section{Conclusions}
We have shown that a 2D-array of tunnel junctions can be used as a 
primary thermometer in the same manner as 1D-arrays demonstrated 
earlier.  2D-arrays have certain advantages over 1D-arrays for this 
application, namely the robustness to damage of individual tunnel 
junctions, higher tolerance to variations in tunnel junction 
properties, lower signal noise and lower resistance which enables 
faster measurement.  This should make them more suitable for 
temperature measurements at temperatures above a few hundred mK where 
the electron heating effect is not a big problem.  At lower 
temperatures 1D-arrays are probably better, since the topology of 
1D-arrays allow big cooling fins to be attached to the metal islands 
of the array, and this will improve the thermalisation of the electron 
system \cite{kaupprb}.

\section*{Acknowledgements}
We thank Jari Kinaret, Jukka Pekola and Juha Kauppinen for fruitful 
discussions.  The Swedish Nanometer Laboratory was used to fabricate 
the samples.  This work was supported financially by the Swedish 
foundation for strategic research (SSF) and by the European Union under 
the TMR program.


\end{document}